\documentclass[aps,prd,superscriptaddress,10pt,showpacs,notitlepage,  
tightnlines,twocolumn,
nofootinbib
]{revtex4-1}
\usepackage{bm}
\usepackage{amsmath,amssymb,amsthm}
\usepackage{latexsym,graphicx,color,subfigure}
\usepackage{enumerate}
\usepackage{amssymb}
\usepackage{hyperref}
\usepackage{mathtools}

	\usepackage{graphicx}
	\graphicspath{ {images/} }

\usepackage{color}

\newcommand{\be}{\begin{equation}}
\newcommand{\ee}{\end{equation}} 
\newcommand{\beq}{\begin{eqnarray}}
\newcommand{\eeq}{\end{eqnarray}}

\newcommand{\p}{\partial}

\newcommand{\bea}{\begin{eqnarray}}
\newcommand{\eea}{\end{eqnarray}}

\def\bra{\langle}
\def\ket{\rangle}

\def\e{\epsilon}

\def\dcfl{\Delta_{\textsc{cfl}}}

\def\bra{\langle}
\def\ket{\rangle}
\def \L{\textsc{l}}
\def \R{\textsc{r}}
\def \B{\textsc{b}}
\def \c{\textsc{c}}
\def \F{\textsc{f}}

\def\cfl{{\rm CFL}}

\usepackage[normalem]{ulem}

\renewcommand\sout{\bgroup \color{blue} \ULdepth=-.5ex \ULset}

\begin{document}
\title{Quark-hadron continuity under rotation: Vortex continuity or boojum?}
\author{Chandrasekhar Chatterjee}
\email{chandra@phys-h.keio.ac.jp}
\affiliation{Department of Physics $\&$ Research and Education Center for Natural Sciences,\\ Keio University,Hiyoshi 4-1-1, Yokohama, Kanagawa 223-8521, Japan}
\author{Muneto Nitta}
\email{nitta(at)phys-h.keio.ac.jp}
\affiliation{Department of Physics $\&$ Research and Education Center for Natural Sciences,\\ Keio University,Hiyoshi 4-1-1, Yokohama, Kanagawa 223-8521, Japan}
\author{Shigehiro Yasui}
\email{yasuis@keio.jp }
\affiliation{Department of Physics $\&$ Research and Education Center for Natural Sciences,\\ Keio University,Hiyoshi 4-1-1, Yokohama, Kanagawa 223-8521, Japan}
\date{\today}
\begin{abstract}
Quark-hadron continuity was proposed as crossover between 
hadronic matter and quark matter without a phase transition, 
based on the matching of the symmetry and excitations in both phases. 
In the limit of a light strange-quark mass,  
it connects hyperon matter and the color-flavor-locked (CFL) phase 
exhibiting color superconductivity. 
Recently, 
it was proposed that this conjecture could
 be generalized in the presence 
of superfluid vortices penetrating both phases~\cite{Alford:2018mqj}, 
and it was
 suggested that 
one hadronic superfluid vortex in hyperon matter 
could be connected to one non-Abelian vortex (color magnetic flux tube) 
in the $\cfl$ phase.
Here, we argue that their proposal is consistent only at large distances;
instead, we show that 
three hadronic superfluid vortices must 
combine with
 three non-Abelian vortices with different colors with 
the total color magnetic fluxes canceled out, 
where the junction is called a colorful boojum.
We rigorously prove this in both a macroscopic theory based on 
the Ginzburg-Landau description 
in which symmetry and excitations match (including vortex cores), 
and a microscopic theory
in which the Aharonov-Bohm phases of quarks around vortices match.

\end{abstract}
\maketitle


\section{Introduction}

The presence or absence of phase transitions is the most important issue to understand phases of matter. 
In the last few decades, a lot of effort was made to understand the phase structure of matter at high density and/or temperature~\cite{Fukushima:2010bq}.
In particular, the region of high density and low temperature 
is relevant for cores of compact stars such as neutron stars,
in which nuclear matter and  quark matter are present. 
The superfluidity of nucleon-nucleon pairing is expected in nuclear matter, 
and 
nuclear matter consisting of hyperons---nuclei containing strange quarks---may be present in high-density regions~\cite{Takatsuka:2000kc} (see Ref.~\cite{Sedrakian:2018ydt} as a recent review).
Quark matter is expected at higher densities; at
asymptotically high density much higher than the strange-quark mass 
the color-flavor-locked ($\cfl$) phase is realized, in which 
three quarks (up ($u$), down ($d$) and strange ($s$)) participate in a diquark pairing, 
exhibiting color superconductivity as well as superfluidity~\cite{Alford:1997zt,Alford:1998mk} (see Refs.~\cite{Alford:2007xm,Rajagopal:2000wf} for a review). 
In addition to superfluid vortices~\cite{Forbes:2001gj,Iida:2002ev},
there are non-Abelian (NA) vortices or color magnetic flux tubes~\cite{Balachandran:2005ev,Nakano:2007dr,Nakano:2008dc,Eto:2009kg,Eto:2009bh,Eto:2009tr,Alford:2016dco} (see Ref.~\cite{Eto:2013hoa} for a review).
The former is dynamically split into three of the latter,
with the total magnetic fluxes canceled out~\cite{Nakano:2007dr,Alford:2016dco}.

The quark-hadron continuity conjecture was proposed  
as a crossover between hadronic matter and 
quark matter, based on the matching of elementary excitations and 
existing global symmetries in both the matter,
in particular, hyperon matter and $\cfl$ phase~\cite{Schafer:1998ef, Alford:1999pa},
as summarized in Table~\ref{table}.
The continuity was further studied in the interior of neutron stars~\cite{Masuda:2012kf,Masuda:2012ed,Baym:2017whm}.
Since neutron stars are rapidly rotating, 
superfluid vortices appear in both
 nuclear matter and quark matter, and
thereby it is natural to extend the quark-hadron continuity in the presence of vortices penetrating both phases of matter~\cite{Alford:2018mqj}.
They defined a continuity of vortices by matching the  Onsager-Feynman circulation of vortices in both phases, 
and suggested a continuity to connect one hadronic vortex to 
one 
NA vortex during the hadron-$\cfl$ crossover. 

In this  paper---by pointing out that the conclusion in Ref.~\cite{Alford:2018mqj} is consistent only for the large-distance behavior of vortices and is not compatible with the symmetry structures of vortex cores---  
we reach the conclusion  that arguably the only possibility left is 
to form a connection of three hadronic vortices in the hyperon matter 
of $\Lambda\Lambda$ condensation  
with three 
NA vortices in the $\cfl$ phase,
with the total color magnetic fluxes canceled out,
forming a colorful boojum~\cite{Cipriani:2012hr}
analogous to a boojum in superfluid helium-3~\cite{Mermin,Volovik}.
We prove this both in a macroscopic theory based on 
the Ginzburg-Landau (GL) description
in which symmetry and excitations match including vortex cores, 
and a microscopic theory 
in which the Aharonov-Bohm (AB) phases of quarks around vortices match.
\begin{table}[bt]
\begin{center}
\begin{tabular}{|c|c|c|}
\hline
 & hadronic phase & CFL phase  \\
\hline
unbroken symmetry & $SU(3)_{\F}$ & $SU(3)_{\c+\F}$ \\
\hline
\# of NG bosons & 8 & 8 \\
\hline
\# of massive vector mesons & 8 & 8 \\
\hline
\# of quasifermions & 8 & 8+1 \\
\hline
 vortex configurations & $\Lambda\Lambda$ &  $ur$, $dg$, $sb$\\
 circulation of one vortex & $2\pi \frac{\nu_{\B}}{2\mu_{\B}}$ & $2\pi \frac{\nu_{\mathrm{q}}}{2\mu_{\mathrm{q}}}$ \\
\hline
\end{tabular}
\end{center}
\caption{The properties in the $\cfl$ phase and in the hadron phase.
}
\label{table}
\end{table}%

\section{Vortex continuity in macroscopic theory}
\label{GL}

The concept of continuity is defined by the continuation of symmetries and elementary excitations in the ground state while going through a crossover.
Now we would like to discuss the concept of continuity in the presence of a general background. For example, the vortices that are present in 
two different phases should be  joined together so that all physical quantities remain smoothly connected and the symmetry structure remains the same through the crossover.

On the other hand, the presence of solitonic objects may break existing unbroken global symmetries present in the ground state. 
Since the condensate eventually reaches its ground-state expectation value (modulo gauge transformations) at large distances, 
the large-distance symmetry structure in general remains
the same as that in the ground state. 
However, this scenario may change inside solitonic objects and  the existing bulk symmetry may be broken spontaneously.  In this case, there appear extra Nambu-Goldstone (NG) zero modes inside the solitons, 
which should be carefully handled during the crossover. 
In other words, to maintain the continuity of solitonic 
objects along with elementary excitations, one should check the symmetry structure everywhere. 

Let us focus our interest on the crossover  
between the hadronic phase and the $\cfl$ phase. 
At high densities, one may expect 
strange quarks to appear
as hyperon states on the hadronic side. 
In general, the first hyperon expected  to appear is $\Lambda$, which is the lightest one with an 
attractive potential in nuclear matter.
For our purposes, here
we consider only flavor-symmetric $\Lambda\Lambda$ pairing in the $^1S_0$ channel.\footnote{We consider the singlet channel as the most attractive one in the $SU(3)_{\F}$ limit. We use the $\Lambda\Lambda$ pairing as an abbreviation of $ -\sqrt{1/8}\,\Lambda\Lambda+\sqrt{3/8}\,\Sigma\Sigma+\sqrt{4/8}\,N\Xi$ pairing for a nucleon ($N$) and $\Lambda$, $\Sigma$, $\Xi$ baryons~\cite{deSwart:1963pdg}.} In this case we may consider the existence of superfluid vortices since $\Lambda\Lambda$ would  break 
$U(1)$ baryon number symmetry and we may express the vortex ansatz  as
\begin{eqnarray}
\label{LL}
\Delta_{\Lambda\Lambda}(r,\theta) = |\Delta_{\Lambda\Lambda}(r)| e^{i\theta},
\end{eqnarray}
where $r$ is the distance from the center of the vortex and $\theta$ is the angle  around the vortex axis.
 The exact nature of the profile function can be derived
from the GL theory of the system, but we are not going to discuss this here. Since the condensate $\Delta_{\Lambda\Lambda}$
is flavor symmetric in this phase, the $SU(3)_\F$ flavor symmetry would be intact everywhere including the vortex cores. The Onsager-Feynman circulation, which is defined as
$
C = \oint \vec v \cdot \mathrm{d} \vec l = \frac{2\pi n}{\mu},
$
(where $n$ and $\mu$ are the winding number and chemical potential of the condensate), can be computed for
a single $\Lambda\Lambda$ vortex as $C_{\Lambda\Lambda} = \frac{2\pi }{2\mu_\B}$,
where $\mu_\B$ is the chemical potential for a single baryon. Here $\vec v$ is the superfluid velocity at large distance from the core of the vortex.

In the $\cfl$ phase, the order parameter is a matrix 
$\Delta_a{}^i = {\Delta_{\L}}_a{}^i   =-{\Delta_{\R}}_a{}^i$ with a color index $a=1,2,3 \,(r,g,b)$ 
and a flavor index $i=1,2,3 \,(u,d,s)$, 
where
${\Delta_{\L}}_a{}^i  \sim  \e_{abc}\e^{\it ijk} {q_\textsc{l}}_b^{\it j} \mathcal{C}{q_\L}_c^k$ and ${\Delta_\R}_a{}^i  \sim  \e_{abc}\e^{ijk} {q_\R}_b^j \mathcal{C}{q_\R}_c^k$. 
The Ginzburg-Landau formulation of the CFL phase has been derived in  Refs.~\cite{Giannakis:2001wz, Iida:2000ha, Iida:2001pg}.
The symmetries in the CFL phase are summarized in Appendix~\ref{CFL-symmetry}.
The order parameter for an Abelian superfluid vortex can be written as~\cite{Forbes:2001gj,Iida:2002ev} 
\begin{eqnarray}
\Delta_{\rm A}(r,\theta) = \dcfl \phi(r)e^{i\theta}\mathbf{1}_3,
\end{eqnarray}
where $\phi(r)$ is a profile function 
vanishing  at the center of the vortex, $\phi(0)= 0$, and eventually reaching the ground state value $\phi(r\to\infty) \rightarrow 1$ at large distances. 
$\dcfl$ is the absolute value of the gap (condensate) at the bulk in 
the CFL phase. 
The Onsager-Feynman circulation of Abelian vortices in the CFL phase is  found to be
$C_{\rm A} = \frac{3\pi}{\mu_\B}$, since the chemical potential of a diquark 
is $\mu_{\rm CFL} = \frac{2\mu_\B}{3}$. 
 So a single $\Lambda\Lambda$ vortex cannot connect continuously
 to a single $U(1)$ vortex in the CFL phase.
 Instead, we may conclude that three $\Lambda\Lambda$ vortices would join to form one $U(1)$ CFL vortex.

\begin{figure}[bt]
\centering
\includegraphics[totalheight=2.6cm]{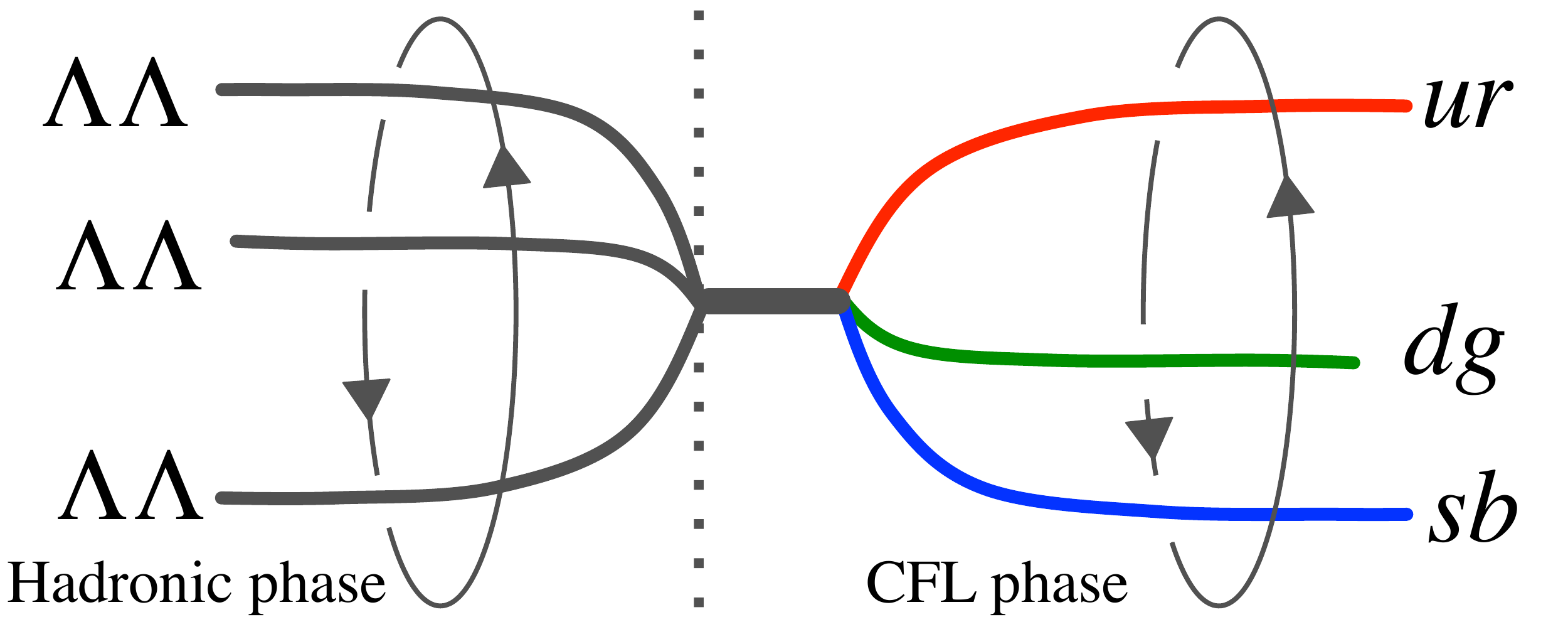}
\caption{A schematic diagram of connection of three $\Lambda\Lambda$ vortices in the hadronic phase with three different 
CFL vortices via single $U(1)$ CFL vortex.}
\label{boojum1}
\end{figure}

Now let us discuss
NA vortices or color magnetic flux tubes.
In this case, the simplest vortex ansatz 
 can be expressed as~\cite{Balachandran:2005ev,Eto:2009kg}
  \begin{eqnarray}
 \label{Ur}
\Delta_{ur}(r, \theta)  &=& 
\dcfl \, {\rm diag}\bigl(f(r)e^{i\theta}, g(r), g(r)\bigr), \label{eq:vortex_ansatz_ur} \\
A^{ur}_i(r) &=&  \frac{1}{3g_s} \frac{\epsilon_{ij} x_j}{r^2} \bigl( 1 - h(r) \bigr) \, {\rm diag}(2, -1,-1),
\end{eqnarray} 
with the gauge coupling constant $g_{s}$.
The profile functions $f(r)$, $g(r)$, and $h(r)$ can be computed numerically with the boundary conditions,
$ f(0) = 0$, $\p_r g(r)|_{r=0} = 0$, $h(0) = 1$,  $\quad f(\infty) = g(\infty) = 1$, and $h(\infty) = 0$~\cite{Eto:2009kg}. 
We call this an up-red ($ur$) vortex since the $ur$ component has a vortex winding. 
We also define two other vortices by changing the
position of the vortex winding ($e^{i\theta}$) from $\Delta_{11}$ to $\Delta_{22}$ and  $\Delta_{33}$, 
which can be called down-green ($dg$) and strange-blue ($sb$) vortices, 
respectively. 
At large distances, the order parameter of these three vortices 
behaves as
\begin{eqnarray} 
 \Delta \sim \dcfl e^{i\theta/3} 
 \exp
 \Biggl(
    - i g_s \int_0^\theta \vec{A}\!\cdot\! \mathrm{d}\vec{l}
 \Biggr)
  \mathbf{1}_{3\times 3}, \label{eq:Delta}
\end{eqnarray}
where 
 $A_i$ is the large-distance configuration of the gauge field corresponding to 
the color flux present inside the vortex core. So it is easy to check that at large distances the $SU(3)_{\c+\F}$ symmetry  remains unbroken. In this case one may derive the superfluid velocity at large distances by replacing
the ordinary derivative with the
covariant derivative in the expression of the current.
The Onsager-Feynman circulation of
NA vortices 
in the $\cfl$ phase is found to be $C_{\rm N\!A} = \frac{\pi}{\mu_\B}$, 
which coincides with the circulation of a single $\Lambda\Lambda$ vortex. 
Therefore one would expect that a single $\Lambda\Lambda$ vortex would be smoothly connected to a single
NA vortex during the crossover~\cite{Alford:2018mqj}. 
Below we show that  this is true only at large distances, and not at short distances near the vortex core. 

First let us consider the symmetry structures in the presence of
NA vortices. According to hadron-quark continuity the unbroken $SU(3)_{\c+\F}$ symmetry
can be smoothly connected to the unbroken flavor symmetry in the hadron phase. So it seems that there would also not be any problem for the continuation of a
NA vortex to 
 a single $\Lambda\Lambda$ vortex. However, the missing point is that the $SU(3)_{\c+\F}$ symmetry present at the bulk is spontaneously broken at the core of an
 NA vortex to $SU(2)\times U(1)$. 
 This generates $\mathbb{C}P^2 \simeq SU(3)/[SU(2)\times U(1)]$ NG modes inside the vortex core~\cite{Nakano:2007dr,Eto:2009bh}. 
 The low-energy effective theory of the $\mathbb{C}P^2$ NG modes 
 was obtained along the vortex line~\cite{Eto:2009bh,Eto:2009tr,Chatterjee:2016tml}.
This helps us to distinguish two different vortices by flavor quantum numbers. 
The three kinds of vortices $ur$, $dg$, and $sb$, where the color part is chosen in a particular gauge for our own convenience, 
lie at three points of 
 the $\mathbb{C}P^2$ 
  moduli space, and they are continuously connected by the flavor symmetry. 
  This can be understood directly from the structure of the order parameters at the center of the vortices.
 We may write the order parameters at the center of the vortices for these three cases as
 \begin{eqnarray}
 \Delta_{ur}(0) &=& c\,\, {\rm diag}(0,1,1), \nonumber \\
 \Delta_{dg}(0) &=& c\,\, {\rm diag}(1,0,1), \nonumber \\
 \Delta_{sb}(0) &=& \, c\,\, {\rm diag}(1,1,0),
\end{eqnarray}
where the constant $c$ can be fixed numerically. 
The flavor symmetry 
$SU(3)_{\c+\F}$ is spontaneously broken by these matrices 
to three different unbroken $SU(2)\times U(1)$ subgroups of 
 $SU(3)_{\c+\F}$. 
 Since  the $SU(3)$ flavor is unbroken in the hadronic vortex, 
 following the symmetry principle of continuity we can say that 
 a single $\Lambda\Lambda$ vortex cannot 
 smoothly transform into any single
 NA vortex.

  We need to have a construction where the $SU(3)$ flavor symmetry is recovered in a vortex core while connecting to the hadronic phase.
  In other words, we have to terminate the ${\mathbb C}P^2$ NG modes.
  This is possible only when three different
  NA vortices combine into
  one $U(1)$ $\cfl$ vortex in whose core the $SU(3)$ flavor symmetry 
  is not broken.\footnote{The reasoning behind the above proposal is related to the fact 
 that a $U(1)$ $\cfl$ vortex is 
 energetically unstable to break into three
 NA vortices~\cite{Nakano:2007dr,Alford:2016dco}.
}
The ${\mathbb C}P^2$ NG modes of the three different kinds of
NA vortices describe fluctuations from the three different points
of the $\mathbb{C}P^2$ moduli space. 
When we combine them,  these NG modes can smoothly
move from one patch to another patch at the junction. 
 As we already discussed,
  one $U(1)$ $\cfl$ vortex can be connected to three $\Lambda\Lambda$  vortices during the hadron-$\cfl$ crossover, 
  and then we reach Fig.~\ref{boojum1}.
  The junction point was called a colorful boojum~\cite{Cipriani:2012hr}, 
  analogous to those in helium-3 superfluids~\cite{Mermin,Volovik}. 

One important point is that 
we do not require the cancellation of color magnetic fluxes at the junction point.
Instead, we only require the termination of the ${\mathbb C}P^2$ NG modes.
The color of a
NA vortex is gauge dependent as emphasized in 
Ref.~\cite{Alford:2018mqj}, but 
the termination of the ${\mathbb C}P^2$ NG modes
implies the cancellation of the color magnetic fluxes 
in our gauge choice.

We comment that the present treatment of the symmetry breaking at the vortex is based on the mean-field approximation. 
The quantum fluctuations 
recover the spontaneously broken symmetry at the vortex core~\cite{Eto:2011mk,Gorsky:2011hd}, 
due to the Coleman-Mermin-Wagner theorem forbidding the existence of spontaneous symmetry breaking or long-range order in 1+1 dimensions.
While NG modes with quadratic dispersion relations in nonrelativistic theories survive at the quantum level~\cite{Nitta:2013wca}, 
NG modes in our case have a linear dispersion relation 
and consequently become massive.

\section{Vortex continuity in microscopic theory}
\label{BDG}
We now prove the same result from a microscopic point of view, 
by requiring the continuity of quark wave functions in the presence of 
vortices penetrating the $\cfl$ and hadronic phases.
More precisely, we achieve the continuity of 
 phases of quarks 
encircling vortices. 

Fermions $\psi$ belonging to the fundamental representation of SU(3) 
acquire an AB phase by 
\begin{eqnarray} 
  \psi \rightarrow \mathrm{P} \, \exp \Biggl({i g_s \oint_{C} \vec{A}\!\cdot\!\mathrm{d}\vec{l}} \Biggr)\psi , \label{eq:AB}
\end{eqnarray} 
for the closed path $C$ in the presence of the gauge field $\vec{A}$.
Notice that the AB phase is the gauge-independent quantity.
Let us consider the case where $C$ encircles a $ur$ NA vortex.
In an appropriate gauge for the SU(3) gauge symmetry, 
the transformation from $\theta$ to $\theta + \alpha$ 
can be written as
\begin{eqnarray}
 \psi
\rightarrow
\,
\mathrm{P} \exp \Biggl(  ig_{s}\int_{\theta}^{\theta+\alpha}\vec{A} \!\cdot\! \mathrm{d}\vec{l} \Biggr)
	 \psi.
\label{eq:AB2}
\end{eqnarray}
For the rotation around the $ur$, $dg$ and $sb$ NA vortices, 
this reduces to
\begin{eqnarray}
 \psi
&\rightarrow&
 \mathrm{diag}\bigl(e^{-i2\alpha/3},e^{i\alpha/3},e^{i\alpha/3}\bigr) \psi,
 \nonumber\\
  \psi 
&\rightarrow&
 \mathrm{diag}\bigl(e^{i\alpha/3},e^{-i2\alpha/3},e^{i\alpha/3}\bigr) \psi
,\nonumber \\
 \psi
&\rightarrow&
 \mathrm{diag}\bigl(e^{i\alpha/3},e^{i\alpha/3},e^{-i2\alpha/3}\bigr) \psi
,
\end{eqnarray}
respectively. 
When the fermion circulates by $2\pi$ rotation ($\alpha=2\pi$), the phase of the fermion changes as 
\begin{eqnarray} 
 \psi \rightarrow \omega \, \psi  \mbox{ with }  \omega=e^{i2\pi/3}
 \label{eq:pure-AB}
\end{eqnarray}
for all cases. 
Heavy quarks 
not participating in the condensation, such as charm quarks,   
feel only this AB phase of $\mathbb{Z}_{3}$~\cite{Cherman:2018jir}.
The independence of the Wilson loops from the species of NA vortices was emphasized in Ref.~\cite{Alford:2018mqj}.
On the other hand, since up, down, and strange quarks participate in the condensation,
they acquire another phase in addition to the AB phase
when they encircle the vortex, as we see below.

In the hadronic phase, we consider the vortex $\Delta_{\Lambda\Lambda}$ in Eq.~(\ref{LL}).
Because $\Delta_{\Lambda\Lambda} \sim \Lambda\Lambda$, the rotation around the vortex from the angle $\theta$ to $\theta+\alpha$
induces the phase for the $\Lambda$ particle component $u_{\B}$ and the $\Lambda$ hole component $v_{\B}$: $(u_{\B},v_{\B}) \to (e^{i\alpha/2}u_{\B},e^{-i\alpha/2}v_{\B})$.
We find that this transformation indeed satisfies the Bogoliubov-de Gennes (BdG) equation in Eq.~(\ref{eq:BdG_equation_Lambda}) in Appendix~\ref{sec:BdG_hadron}.
A complete encircling ($\alpha=2\pi$) yields 
a phase factor $\pm \pi$ for both the particle and hole components.
This is not an AB phase, because 
there is no gauge field in the $\Lambda\Lambda$ vortex.
This is induced purely by the angle dependence of the vortex, 
and so 
we call this quantity a vortex phase of 
the $\Lambda$ baryon  around the vortex.
This phase factor forms a $\mathbb{Z}_{2}$ group.

Let us understand this vortex phase factor at the quark level.
Since the $u,d,s$ quarks are confined inside the $\Lambda$ baryon in a symmetric way, each quark's quasiparticles $q_{\alpha i}$ in $\Delta_{\Lambda\Lambda}$ should have vortex phases $Q_{\Lambda\Lambda}(q)$ 
given by
\begin{eqnarray}
 Q_{\Lambda\Lambda}
=
\frac{\pi}{3}
\left(
\begin{array}{ccc}
 +1 & +1 & +1 \\
 +1 & +1 & +1 \\
 +1 & +1 & +1
\end{array}
\right)
 \; {\rm for}\;
(q)_{a i}
\!=\!
\left(
\begin{array}{ccc}
 u_{r} & d_{r} & s_{r} \\
 u_{g} & d_{g} & s_{g} \\
 u_{b} & d_{b} & s_{b}
\end{array}
\right).
\label{quark_charge}\nonumber\\
\end{eqnarray}
The matrix $Q_{\Lambda\Lambda}$ is defined so that the component $(Q_{\Lambda\Lambda})_{ai}$ represents the vortex phase for the quark $q_{a}^{i}$; $q_{a}^{i}\rightarrow e^{i(Q_{\Lambda\Lambda})_{ai}}q_{a}^{i}$ for a rotation by $2\pi$.
In fact, $Q_{\Lambda\Lambda}$ is obtained by setting $\alpha=2\pi$ for the vortex phase change $e^{i\alpha/6}$ of quarks in the $\Lambda$ baryon for an angular transformation from $\theta$ to $\theta + \alpha$.
Thus, the vortex phase of the $u,d,s$ quarks in the hadronic phase forms a $\mathbb{Z}_{6}$ group.
Notice that the phase should be independent of flavor and color.
Here, we indicated the vortex phases of the particle components, 
but the hole components simply have the opposite sign: 
the hole components obtain the vortex phase $-\pi/3$.
The heavy quarks not participating in the condensation 
receive no vortex phases.
The question is then how the phases of quarks can be connected smoothly from the hadronic phase to the $\cfl$ phase.

Before investigating the vortex phases and AB phases of quarks in the $\cfl$ phase, we recall the simpler case of a single-component Dirac fermion ($\psi$) in the BdG equation which has a vortex phase $\pi$ because of the phase $\alpha/2$, as in $\psi \rightarrow e^{i\alpha/2}\psi$ for the rotation $\theta \rightarrow \theta + \alpha$, as shown in Appendix~\ref{sec:BdG_single_Dirac}.

Now, let us investigate the vortex phases and AB phases for $u,d,s$ 
quarks 
around NA vortices, 
where the quarks couple to the gap profile functions differently.
We remember that the color and flavor structure in the gap is $\Delta_{ai} \sim  \epsilon_{abc}\epsilon^{ijk} q_{b}^{j}{\cal C}q_{c}^{k}$ (which indicates $a,b,c$ for color and $i,j,k$ for flavor).
In general, the phases of the $u,d,s$ quarks in the NA vortex change as
\begin{eqnarray}
 q
\rightarrow
e^{i\alpha/6} \,
\mathrm{P} \exp \Biggl(  ig_{s}\int_{\theta}^{\theta+\alpha}\vec{A} \!\cdot\! \mathrm{d}\vec{l} \Biggr)
 q ,
\label{eq:phase_change_quark_ur_0}
\end{eqnarray}
for a rotation by $\alpha$; $\theta \rightarrow \theta+\alpha$.
The important point is that the vortex phase factor $e^{i\alpha/6}$,
which is a contribution from 
the global transformation of $U(1)_{\rm B}$
and is missing in the case of heavy quarks in Eq.~(\ref{eq:AB2}), 
 must be present to be consistent with the condensation with 
 the vortex; since a diquark $\Delta$ receives $e^{i\alpha/3}$, 
 the $u,d,s$ quarks receive $e^{i\alpha/6}$ (see Eq.~(\ref{eq:Delta})).
The remaining part 
is the usual AB phase coming from the gauge symmetry 
where $\vec{A}$ is the color gauge field around a NA vortex 
[note the difference in signs between 
Eqs.~(\ref{eq:phase_change_quark_ur_0}) and (\ref{eq:Delta})].
We call this total  phase introduced in Eq.~(\ref{eq:phase_change_quark_ur_0}) a generalized AB phase.

Let us consider the cases of the $ur$, $dg$ and $sb$ NA vortices 
for illustration. 
In the appropriate gauge, the above transformation is reduced to
\begin{eqnarray}
 q
&\rightarrow&
e^{i\alpha/6} \,
\mathrm{diag}
\bigl(e^{-i2\alpha/3},e^{i\alpha/3},e^{i\alpha/3}\bigr)
 q
\nonumber \\
&=&
\mathrm{diag}
\bigl(e^{-i\alpha/2},e^{i\alpha/2},e^{i\alpha/2}\bigr)
 q, 
 \nonumber\\
 q
&\rightarrow&
e^{i\alpha/6} \,
\mathrm{diag}
\bigl(e^{i\alpha/3},e^{-i2\alpha/3},e^{i\alpha/3}\bigr)
 q
\nonumber \\
&=&
\mathrm{diag}
\bigl(e^{i\alpha/2},e^{-i\alpha/2},e^{i\alpha/2}\bigr)
 q,
 \nonumber\\
 q
&\rightarrow&
e^{i\alpha/6} \,
\mathrm{diag}
\bigl(e^{i\alpha/3},e^{i\alpha/3},e^{-i2\alpha/3}\bigr)
 q
 \nonumber \\
&=&
\mathrm{diag}
\bigl(e^{i\alpha/2},e^{i\alpha/2},e^{-i\alpha/2}\bigr)
 q,
\label{eq:phase_change_quark}
\end{eqnarray}
for the $ur$, $dg$, and $sb$ NA vortices, respectively. 
We can easily confirm that these transformations 
satisfy the BdG equation for the quarks in the $ur$, $dg$, and $sb$ NA vortices.
The explicit form of the BdG equation is presented in Eq.~(\ref{eq:CFL_eve}) in Appendix~\ref{sec:BdG_NA_quark}. 
Thus, by setting $\alpha=2\pi$, we obtain the common value for the generalized AB phases $Q_{ur}(q)$, $Q_{dg}(q)$, $Q_{sb}(q)$ of $u,d,s$ quarks around the $ur$, $dg$, and $sb$ vortices, respectively, as
\begin{eqnarray}
Q_{ur}
&=&
\pi
\left(
\begin{array}{ccc}
 -1 & -1 & -1 \\
 +1 & +1 & +1 \\
 +1 & +1 & +1
\end{array}
\right),\,
Q_{dg}
=
\pi
\left(
\begin{array}{ccc}
 +1 & +1 & +1 \\
 -1 & -1 & -1 \\
 +1 & +1 & +1
\end{array}
\right),
\nonumber \\
Q_{sb}
&=&
\pi
\left(
\begin{array}{ccc}
 +1 & +1 & +1 \\
 +1 & +1 & +1 \\
 -1 & -1 & -1
\end{array}
\right).
\label{quark_charge_Sb}
\end{eqnarray}
They give the same phases because $e^{\pm i \pi}=-1$.
Therefore, the generalized AB phases of any of the $u,d,s$ quarks around 
any NA vortex forms a $\mathbb{Z}_{2}$ group.
Hereafter, we write $Q_{\rm NA}=\{Q_{ur},Q_{dg},Q_{sb}\}$.

Now we consider the connection of the vortices in the hadronic and $\cfl$ phases,  by requiring the continuity of quark wave functions, that is, the matching of generalized AB phases (including vortex phases) in both phases.
The generalized AB phase  $\pi/3$
 of quarks around a $\Lambda\Lambda$ vortex 
in Eq.~(\ref{quark_charge}) 
is apparently different from the phase $\pi$ of quarks in any single
NA vortex (either $ur$, $dg$, or $sb$) in Eq.~(\ref{quark_charge_Sb}): 
\begin{eqnarray}
  e^{i(Q_{\Lambda\Lambda})_{ai}} \neq e^{i(Q_{\rm NA})_{ai}}.
\end{eqnarray}
This mismatch can also be understood by the differences between the groups for 
$u,d,s$ quarks: the $\mathbb{Z}_{6}$ group in the hadronic phase and the $\mathbb{Z}_{2}$ group in the CFL phase.
Therefore, one 
NA vortex cannot be connected to one $\Lambda\Lambda$ vortex without discarding the continuity at the quark level, although such a connection could be consistent only at large distance scales in the GL equation~\cite{Alford:2018mqj}.

To achieve a smooth connection with the generalized AB phases in 
Eq.~(\ref{quark_charge_Sb}), 
we may consider the matching of the generalized AB phase: $e^{i3(Q_{\Lambda \Lambda})_{ai}}=e^{i(Q_{ur})_{ai}}$, $e^{i3(Q_{\Lambda \Lambda})_{ai}}=e^{i(Q_{dg})_{ai}}$, or $e^{i3(Q_{\Lambda \Lambda})_{ai}}=e^{i(Q_{sb})_{ai}}$.
However, those cases violate the circulation matching, and hence they should be discarded (cf.~Sec.~\ref{GL}).

Alternatively, we consider the case where the quark encircles a {\it bundle} of three NA
 vortices simultaneously.
We notice that all of the quarks acquire the generalized AB phase $Q_{\rm NA}$ irrespective of the flavor and color components.
For example, the $ur$ quark acquires the phase $Q_{\rm NA}$ for the path around 
each vortex, and hence it acquires $3Q_{\rm NA}$ in total.
The generalized AB phases in the path encircling the three NA vortices simultaneously are equal to the sum of the generalized AB phases in the paths encircling each of them: $Q_{3 {\rm NA}}=3Q_{\rm NA}$.
As a result, the quarks with any flavor and color acquire a common charge, which is equal to the generalized AB phase in the presence of {\it three} $\Lambda\Lambda$ vortices,
\begin{eqnarray}
 e^{i(Q_{3 {\rm NA}})_{ai}} (= e^{i3(Q_{\rm NA})_{ai}}) 
=e^{i3(Q_{\Lambda\Lambda})_{ai}}. \label{eq:QQQ=3Q}
\end{eqnarray}
Therefore, the continuity of the generalized AB phases of quarks is allowed only when the bundle of three NA vortices is connected to the bundle of three $\Lambda\Lambda$ vortices.
Eq.~(\ref{eq:QQQ=3Q}) does not imply species of NA vortices. On the other hand, one Abelian vortex is unstable to decay into three NA vortices with different color magnetic fluxes with total color flux canceled out~\cite{Nakano:2007dr,Alford:2016dco}. In fact, we have a more precise relation
\begin{eqnarray}
3Q_{\Lambda\Lambda}=Q_{ur}+Q_{dg}+Q_{sb}.
\end{eqnarray}
It is interesting to point out that this relation holds without an exponential function.

We prove that the three NA vortices and the three $\Lambda\Lambda$ vortices meet at {\it one} point in transverse directions (see Fig.~\ref{boojum1}).
First of all, we notice that the quarks can take an arbitrary path. One may think of a path that does not necessarily encircle all of the NA vortices   when those vortices are separated in space.
However, such a path precludes the continuity of the quark wave functions between the $\cfl$ and hadronic phases.
Therefore, only the paths that simultaneously encircle the three NA vortices should be allowed to exist: the three NA vortices meet at one point.
There, they are connected to a $U(1)_{\rm B}$ Abelian vortex (a boojum), as shown in Fig.~\ref{boojum1}.
In summary, the continuity of the quark wave function induces that the bundle of the three 
vortices and the three $\Lambda\Lambda$ vortices are connected via the $U(1)$ vortex.

\begin{widetext}
\begin{center}
\begin{table}[bt]
\begin{center}
\begin{tabular}{|c|c|c|c|c|}
\hline
   \multicolumn{2}{|c|}{} & hadronic phase & \multicolumn{2}{|c|}{CFL phase} \\
\hline
 \multicolumn{2}{|c|}{vortex} & $\Lambda\Lambda$ vortex &  Abelian vortex & NA vortex \\
\hline \hline
heavy quarks & AB phase & $\mathbf{1}$ & $\mathbf{1}$ & $\mathbb{Z}_{3}$ \\
\hline
 $u,d,s$ & AB phase & $\mathbf{1}$ & $\mathbf{1}$ & $\mathbb{Z}_{3}$ \\
 & generalized AB phase & $\mathbb{Z}_{6}$ & $\mathbb{Z}_{2}$ & $\mathbb{Z}_{2}$ \\
\hline
\end{tabular}
\end{center}
\caption{The groups for the (generalized) AB phases 
around a $\Lambda\Lambda$ vortex in the hadronic phase and 
a NA vortex in the CFL phase.}
\label{table2}
\end{table}%
\end{center}
\end{widetext}

In Table~\ref{table2}, we summarize the (generalized) 
AB phases for heavy quarks and $u,d,s$ quarks.
The AB phases in both a $\Lambda\Lambda$ vortex in the hadronic phase and an Abelian vortex in the CFL phase are the trivial group ($\mathbf{1}$) because these vortices have no color magnetic fluxes. 
From this table, we immediately find that three $\Lambda\Lambda$ vortices and three NA vortices should be connected in order to achieve the matching of the generalized AB phases of the $u,d,s$ quarks between the hadronic and CFL phases.
This is explained as follows.
Let us consider the AB phase for heavy quarks, 
and the generalized AB phase for $u,d,s$ quarks in each vortex.
First, the bundle of three $\Lambda\Lambda$ vortices can be connected to one Abelian vortex, because three $\mathbb{Z}_{6}$'s in three $\Lambda\Lambda$ vortices become equivalent to $\mathbb{Z}_{2}$ in an Abelian vortex.
Second, the bundle of three NA vortices can be connected to one Abelian vortex, because three $\mathbb{Z}_{3}$'s in three NA vortices become equivalent to 
$\mathbf{1}$ for the Abelian vortex.
Thus, three $\Lambda\Lambda$ vortices should be connected to three NA vortices through one  Abelian vortex (a boojum).
Therefore, we have proven (i) a three-to-three correspondence between $\Lambda\Lambda$ vortices and NA vortices, and (ii) the existence of the boojum.

\section{Summary and Discussion}
\label{conclusion}
In this paper we have  discussed the continuity of vortices during the crossover between the hadronic and CFL phases. 
By using macroscopic (GL) and microscopic (quark) descriptions, 
we have proved that  three $\Lambda\Lambda$ vortices in the hadronic phase must combine and transform into three different 
NA vortices ($ur$, $dg$, $sb$)
in the $\cfl$ phase to maintain a smooth connection 
[see Eq.~(\ref{eq:QQQ=3Q})]. 
The colorful boojum is inevitable for quark-hadron continuity.

We have ignored (strange) quark masses and electromagnetic interactions, whose effects on an 
NA vortex were 
investigated in Refs.~\cite{Eto:2009tr} and~\cite{Vinci:2012mc,Hirono:2012ki,Chatterjee:2015lbf}, respectively.
We should take them into account for more realistic situations.
One question is whether 
fermion zero modes in a vortex core in the CFL phase~\cite{Yasui:2010yw,Fujiwara:2011za} 
and their braiding statistics~\cite{Yasui:2010yh,Hirono:2012ad}
have a continuous transition to the hadron phase.
Another interesting question is what role a confined monopole in the CFL phase~\cite{Eto:2011mk,Gorsky:2011hd} plays 
for quark-hadron duality.
It is also interesting to study how vortex lattices~\cite{Kobayashi:2013axa} are connected during continuity.
Finally, it will be important to study impacts of 
the presence of vortex junctions (boojums) on dynamics of 
neutron stars.

\section*{Acknowledgments}
We would like to thank Motoi Tachibana for discussions.
This work is supported by the Ministry of Education, Culture, Sports, Science and Technology (MEXT)-Supported Program for the Strategic Research Foundation at Private Universities ``Topological Science" (Grant No. S1511006). 
C.~C. acknowledges support as an International Research Fellow of the Japan Society for the Promotion of Science (JSPS) (Grant No: 16F16322). 
This work is also supported in part by 
JSPS Grant-in-Aid for Scientific Research (KAKENHI Grant No. 16H03984 (M.~N.), No.~18H01217 (M.~N.), No.~17K05435 (S.~Y.)), and also by MEXT KAKENHI Grant-in-Aid for Scientific Research on Innovative Areas ``Topological Materials Science'' No.~15H05855 (M.~N.).

\appendix

\section{symmetries of the $\cfl$ Phase}
\label{CFL-symmetry}
We summarize the symmetries of the CFL phase.
The color-flavor-locked phase can be expected when the density becomes asymptotically high. 
The order parameters in the $\cfl$ phase  are defined  by the diquark condensates (close to the critical temperature $T_{\rm c}$) as
${\Delta_{\L}}_a^i  \sim  \e_{abc}\e^{ijk} {q_\textsc{l}}_b^j \mathcal{C}{q_\L}_c^k$ and $\quad {\Delta_\R}_a^i  \sim  \e_{ abc}\e^{ijk} {q_\R}_b^j \mathcal{C}{q_\R}_c^k$,
 where $q_{\L / \R}$ are left-/right-handed quarks carrying 
 fundamental color indices ${a, b, c}$ ($SU(3)_{\c}$) and  fundamental flavor indices $i,j,k$ ($SU(3)_{\L/\R}$). 
  The chiral symmetry is spontaneously broken in the ground state  
  $\Delta_\L = - \Delta_\R \equiv \Delta$. 
  The order parameter 
  $\Delta$ transforms as
  $\Delta' = e^{i\theta_\B} U_\c^* \Delta U_\F^{\dagger}$, 
  $e^{i\theta_\B} \in U(1)_\B$,
  $ U_\c \in SU(3)_\c$, and
  $ U_\F \in SU(3)_\F$.
After subtracting the redundant discrete symmetries, the actual symmetry group is given by
$G  =  \scriptstyle
    \dfrac{\scriptstyle SU(3)_{\c} \times SU(3)_{\F} \times U(1)_{\B}}
   {\scriptstyle\mathbb{Z}_3 \times \mathbb{Z}_3}
\label{eq:sym_G}$. 
In the ground state the full symmetry group $G$ is spontaneously broken down 
to  $\displaystyle{{H} \simeq \,\scriptstyle SU(3)_{\c+\F}/ \scriptstyle\mathbb{Z}_3}\,$
 and the order parameter is defined as
$\bra \Delta \ket = \dcfl {\bf 1_3}$, where  
$\dcfl $ depends on the GL parameters~\cite{Giannakis:2001wz, Iida:2000ha, Iida:2001pg}. 
The elements of the unbroken group $SU(3)_{\c+\F}$ are defined by the relation $U_\c^* = U_\F = U \in SU(3)_{\c+\F} $.
The quarks transform as adjoint field under $SU(3)_{\c+\F}$ as $q' = U q U^\dagger$.
The existence of stable vortices  can be confirmed by a nontrivial
first homotopy group of the order parameter space $\pi_1 (G/H) \simeq \mathbb Z$.

\section{BdG equation}
\subsection{Hadronic matter}
\label{sec:BdG_hadron}
The BdG equation in the $\Lambda\Lambda$ vortex ($\Delta_{\Lambda\Lambda}$ in Eq.~(\ref{LL})) can be written as
\begin{eqnarray}
\left(
\begin{array}{cc}
 -\frac{\vec{\nabla}^{2}}{2m_{\B}} - \mu_{\B} & e^{i\theta}|\Delta_{\Lambda\Lambda}| \\
 e^{-i\theta}|\Delta_{\Lambda\Lambda}| & \frac{\vec{\nabla}^{2}}{2m_{\B}}+ \mu_{\B}
\end{array}
\right)
\left(
\begin{array}{c}
 u_{\B} \\
 v_{\B}
\end{array}
\right)
=
{\cal E}
\left(
\begin{array}{c}
 u_{\B} \\
 v_{\B}
\end{array}
\right)
\label{eq:BdG_equation_Lambda}
\end{eqnarray}
for the $\Lambda$ baryon (with particle component $u_{\B}$ and hole component $v_{\B}$) in the Nambu-Gor'kov formalism.
Here, $m_{\mathrm{B}}$ is the $\Lambda$ baryon mass and $|\Delta|$ is the profile function of the vortex.
The transformation $(u_{\B},v_{\B}) \to (e^{i\alpha/2}u_{\B},e^{-i\alpha/2}v_{\B})$ satisfies the BdG equation.

\subsection{Single-component Dirac fermion}
\label{sec:BdG_single_Dirac}
We consider single-component (massless) Dirac fermion in the presence of a vortex with winding number $1$.
The explicit form of the BdG equation is
\begin{eqnarray}
\left(
\begin{array}{cc}
 -i\gamma_{0} \vec{\gamma} \!\cdot\! \vec{\nabla} \!-\! \mu & e^{i\theta}|\Delta| \gamma_{0} \gamma_{5} \\
-e^{-i\theta}|\Delta| \gamma_{0} \gamma_{5} & -i\gamma_{0} \vec{\gamma} \!\cdot\! \vec{\nabla} \!+\! \mu
\end{array}
\right)
\!\!
\left(
\begin{array}{c}
 \!u\! \\
 \!v\!
\end{array}
\right)
\!\!=\!
{\cal E}
\!
\left(
\begin{array}{c}
 \!u\! \\
 \!v\!
\end{array}
\right),
\end{eqnarray}
with particle component $u$ and hole component $v$ in the Nambu-Gor'kov representation.
Here $\mu$ is the chemical potential.
The rotation of the quark around the vortex changes $\theta$ to $\theta + \alpha$.
This is compensated by the phase rotations for $u$ and $v$, to maintain the above equation, by changing $(u,v)$ to $(e^{i\alpha/2}u,e^{-i\alpha/2}v)$.
Therefore, the particle (hole) 
in the presence of the vortex has a vortex phase $\pm \pi$ for the rotation $\alpha=2\pi$.

\subsection{Quarks in an NA vortex}
\label{sec:BdG_NA_quark}
The BdG equation 
${\cal H}\Psi = {\cal E}\Psi$ in the presence of a $sb$
 vortex is given by~\cite{Yasui:2010yw,Fujiwara:2011za}  
 (see also Ref.~\cite{Sadzikowski:2002in})
\begin{widetext}
\begin{eqnarray}
\left(
\begin{array}{ccccccccc}
 \hat{\cal H}_0 & \hat{\Delta}_{1} & \hat{\Delta}_{0} & 0 & 0 & 0 & 0 & 0 & 0 \\
 \hat{\Delta}_{1} & \hat{\cal H}_0 & \hat{\Delta}_{0} & 0 & 0 & 0 & 0 & 0 & 0 \\
 \hat{\Delta}_{0} & \hat{\Delta}_{0} & \hat{\cal H}_0 & 0 & 0 & 0 & 0 & 0 & 0 \\
 0 & 0 & 0 & \hat{\cal H}_0 & -\hat{\Delta}_{1} & 0 & 0 & 0 & 0 \\
 0 & 0 & 0 & -\hat{\Delta}_{1} & \hat{\cal H}_0  & 0 & 0 & 0 & 0 \\
 0 & 0 & 0 & 0 & 0 & \hat{\cal H}_0 & -\hat{\Delta}_{0} & 0 & 0 \\
 0 & 0 & 0 & 0 & 0 & -\hat{\Delta}_{0} & \hat{\cal H}_0 & 0 & 0 \\
 0 & 0 & 0 & 0 & 0 & 0 & 0 & \hat{\cal H}_0 & -\hat{\Delta}_{0} \\
 0 & 0 & 0 & 0 & 0 & 0 & 0 & -\hat{\Delta}_{0} & \hat{\cal H}_0
\end{array}
\right)
\left(
\begin{array}{c}
 u_r \\
 d_g \\
 s_b \\
 d_r \\
 u_g \\
 s_r \\
 u_b \\
 s_g \\
 d_b
\end{array}
\right)
= {\cal E}
\left(
\begin{array}{c}
 u_r \\
 d_g \\
 s_b \\
 d_r \\
 u_g \\
 s_r \\
 u_b \\
 s_g \\
 d_b
\end{array}
\right),
\label{eq:CFL_eve} 
\end{eqnarray}
\end{widetext}
where we have used the notation e.g., 
$u_{r}$ 
 in the Nambu-Gor'kov representation.\footnote{The BdG equation was used to find a Majorana fermion zero mode in an
 NA vortex in Refs.~\cite{Yasui:2010yw,Fujiwara:2011za}, and it was applied to the non-Abelian statistics of exchanging multiple 
 NA vortices~\cite{Yasui:2010yh,Yasui:2011gk,Hirono:2012ad,Yasui:2012zb}. The coupling of the Majorana fermion zero modes and the $\mathbb{C}P^2$ NG modes was obtained in Ref.~\cite{Chatterjee:2016ykq}.}
We define
$\hat{\cal H}_{0} =
\mathrm{diag}
\bigl(
  -i\gamma_{0} \vec{\gamma} \cdot \vec{\nabla} - \mu_{q},
  -i\gamma_{0} \vec{\gamma} \cdot \vec{\nabla} + \mu_{q}
\bigr)$
and
\begin{eqnarray}
\hat{\Delta}_{i} =
\left(
\begin{array}{cc}
 0 & \Delta_{i} \gamma_{0} \gamma_{5} \\
 -\Delta_{i}^{\dag} \gamma_{0} \gamma_{5} & 0
\end{array}
\right) \hspace{1em} (i=0,1),
\end{eqnarray} 
where $\Delta_{1}(r,\theta) = |\Delta_{1}(r)|\, {\rm e}^{i\theta}$ 
corresponds to the vortex configuration with winding number 1, 
and $\Delta_{0}(r)$ does not have a winding number.

\end{document}